\documentclass{jfm}

\usepackage{graphicx}
\usepackage{natbib}
\usepackage{textcomp}
\usepackage{mathtools}
\usepackage{stmaryrd} 
\usepackage{makecell}
\usepackage{booktabs}
\usepackage{leftidx}
\usepackage{tikz}
\usetikzlibrary{arrows,calc,positioning}
\usepackage{pgfplots}
\pgfplotsset{compat=1.3}
\usepackage{pifont}
\usepackage{xspace}
\usepackage{microtype}
\usepackage{pdflscape}
\ifCUPmtlplainloaded \else
  \checkfont{eurm10}
  \iffontfound
    \IfFileExists{upmath.sty}
      {\typeout{^^JFound AMS Euler Roman fonts on the system,
                   using the 'upmath' package.^^J}%
       \usepackage{upmath}}
      {\typeout{^^JFound AMS Euler Roman fonts on the system, but you
                   dont seem to have the}%
       \typeout{'upmath' package installed. JFM.cls can take advantage
                 of these fonts,^^Jif you use 'upmath' package.^^J}%
      }
  \else
  \fi
\fi

\ifCUPmtlplainloaded \else
  \checkfont{msam10}
  \iffontfound
    \IfFileExists{amssymb.sty}
      {\typeout{^^JFound AMS Symbol fonts on the system, using the
                'amssymb' package.^^J}%
       \usepackage{amssymb}%
         \let\leq=\leqslant
         
      }{}
  \fi
\fi

\ifCUPmtlplainloaded \else
  \IfFileExists{amsbsy.sty}
    {\typeout{^^JFound the 'amsbsy' package on the system, using it.^^J}%
     \usepackage{amsbsy}}
    {\providecommand\boldsymbol[1]{\mbox{\boldmath $##1$}}}
\fi

\newcommand\Ca{\mbox{\textit{Ca}}}  

\newsavebox{\astrutbox}
\sbox{\astrutbox}{\rule[-5pt]{0pt}{20pt}}

\newcommand{\uohne}{\boldsymbol{\mathfrak{u}}}
\newcommand{\umit}{\boldsymbol{\hat{\mathfrak{u}}}}
\newcommand{\Tohne}{\boldsymbol{\mathfrak{T}}}
\newcommand{\Tmit}{\boldsymbol{\hat{\mathfrak{T}}}}
\newcommand{\bild}[1]{figure~\ref{#1}}

\newcommand{\gleich}[1]{equation~\eqref{#1}}

\newcommand{\definiert}{\mathrel{\mathop:}=}

\newcommand{\vek}[1]{\boldsymbol{#1}}

\newcommand{\ord}[1]{^{(#1)}\xspace}

\newcommand{\landaueps}[1]{\mathcal{O}(\varepsilon^{#1})}

\newcommand{\kYn}{\leftidx{_k}{Y_n\ord1}}
\newcommand{\kXn}{\leftidx{_k}{X_n\ord1}}
\newcommand{\kZn}{\leftidx{_k}{Z_n\ord1}}
\newcommand{\kchin}{\leftidx{_k}{\chi}_{-n-1}\ord1}
\newcommand{\kphin}{\leftidx{_k}{\phi}_{-n-1}\ord1}
\newcommand{\kpn}{\leftidx{_k}{p}_{-n-1}\ord1}

\makeatletter

\newcommand{\Rmnum}[1]{{\expandafter\@slowromancap\romannumeral #1@}}
\makeatother

\title[]{Drag and diffusion coefficients of a spherical particle attached to a fluid interface}

\author[Aaron D\"orr and Steffen Hardt]%
{Aaron D\"orr and Steffen Hardt}

\affiliation{Institute for Nano- and Microfluidics, Center of Smart Interfaces, Technische Universit\"at Darmstadt, Alarich-Weiss-Stra\ss e 10, 64287 Darmstadt, Germany}

\date{\today}

\begin{document}

\maketitle
\begin{abstract}
Explicit analytical expressions for the drag and diffusion coefficients of a spherical particle attached to the interface between two immiscible fluids are constructed for the case of a small viscosity ratio between the fluid phases. The model is designed to explicitly account for the dependence on the contact angle between the two fluids and the solid surface. The Lorentz reciprocal theorem is applied in the context of a geometric perturbation approach, which is based on the deviation of the contact angle from a~90\textdegree{}-value. By testing the model against experimental and numerical data from the literature, good agreement is found within the entire range of contact angles below~90\textdegree. As an advantage of the method reported, the drag and diffusion coefficients can be calculated up to second order in the perturbation parameter, while it is sufficient to know the velocity and pressure fields only up to first order. Extensions to other particle shapes with known velocity and pressure fields are straightforward.
\end{abstract}

%
%
\section{Introduction}
The diffusive behavior of colloidal particles is drastically altered compared to diffusion in a bulk fluid when the particles are affected by the presence of an interface between two immiscible fluids. The motion of particles attached to a fluid interface occurs predominantly parallel to the interface, but may also involve temporary particle detachment, as can be concluded from experiments \citep{Walder2010,Sriram2012}. The phenomenon of two-dimensional interfacial diffusion is not yet fully understood, which is reflected in the variety of experimental results as well as related theoretical models. For example, \citet{Peng2009} and \citet{Chen2008} have studied the influence of the surface concentration of diffusing particles on the diffusion coefficient. In the limit of infinite dilution, the measured diffusion coefficient is found to be very close to the bulk value in one of the fluid phases. The authors of both publications explain the data by assuming the interface as incompressible, although there is no reason to assume contamination of the interface \citep{Peng2009}. As another example, experimentalists have occasionally found the size dependence of the diffusion coefficient to differ from the inverse of the particle radius,~$a^{-1}$ \citep{Du2012,Wang2011}, at variance with the modified Stokes-Einstein relation \citep{Brenner1978}
\begin{equation}
D=\frac{kT}{6\pi\mu_1 a f(\Theta,\mu_2/\mu_1)}.
\label{eq:StokesEinstein}
\end{equation}
By the index~1 we denote the fluid with the higher viscosity~$\mu_1$, while index~2 denotes the fluid having the lower viscosity. The three-phase contact angle~$\Theta$ is measured in fluid~1. In \gleich{eq:StokesEinstein}, the function~$f$ specifies the deviation of the drag force from the Stokes drag of a spherical particle suspended in the bulk of fluid~1. In terms of the drag force~$\vek F_D$ acting on the attached particle, $f$ is thus defined by
\begin{equation}
\vek F_D=-6\pi\mu_1af\vek U,
\label{eq:fdef}
\end{equation}
where~$\vek U$ denotes the particle velocity relative to the undisturbed fluids. Even when the modified Stokes-Einstein relation~\eqref{eq:StokesEinstein} is valid, as we shall assume in this study, the functional form of the drag coefficient~$f$ of a translating interfacial particle is not known. A variety of theoretical and experimental studies deal with the drag coefficient of particles attached to fluid-fluid interfaces \citep{Fulford1986,O'Neill1986,Petkov1995,Danov1995,Danov1998,Cichocki2004,Fischer2006,Pozrikidis2007,Ally2010,Bawzdziewicz2010}. However, the corresponding theoretical models mostly rely on numerical methods. With this work, we intend to contribute to the field by providing an explicit analytical expression for the drag coefficient of a spherical particle attached to a pure interface between two fluids of highly different viscosity. According to the Stokes-Einstein relation~\eqref{eq:StokesEinstein}, the diffusion coefficient~$D$ directly follows from the drag coefficient~$f$. Our analysis aims at the latter, which can be studied by means of low Reynolds number hydrodynamics. The main feature of the new model is to account for the dependence of the drag and diffusion coefficients on the three-phase contact angle.
\section{Series Expansion of the Flow Field}
\begin{figure}
\centering%
\begin{tikzpicture}[every node/.style={inner sep=0pt},x=1cm,y=1cm]
\node[name=a] at (0,0) {\includegraphics[scale=0.02]{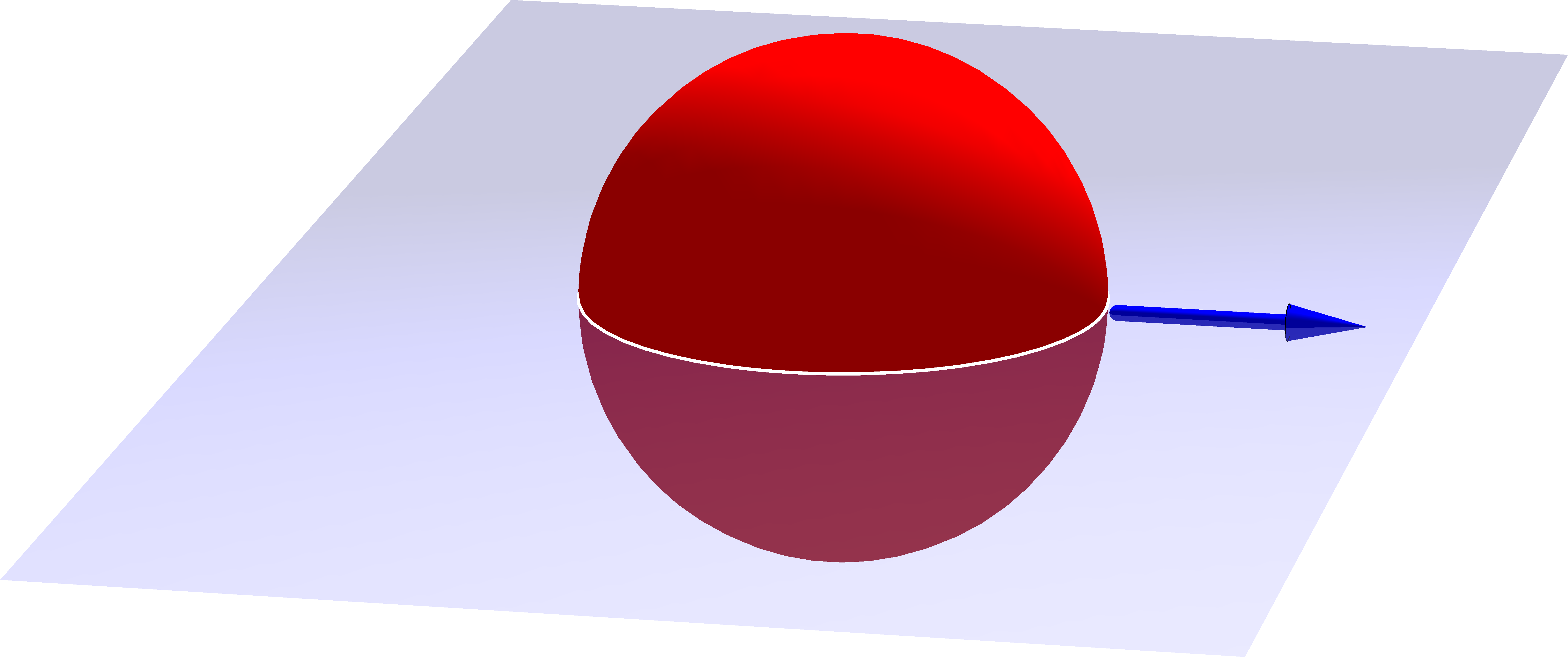}};
\node[above=0.5cm of a.north] {};
\node[name=b,right=of a,yshift=-0.13cm] {\includegraphics[scale=0.02]{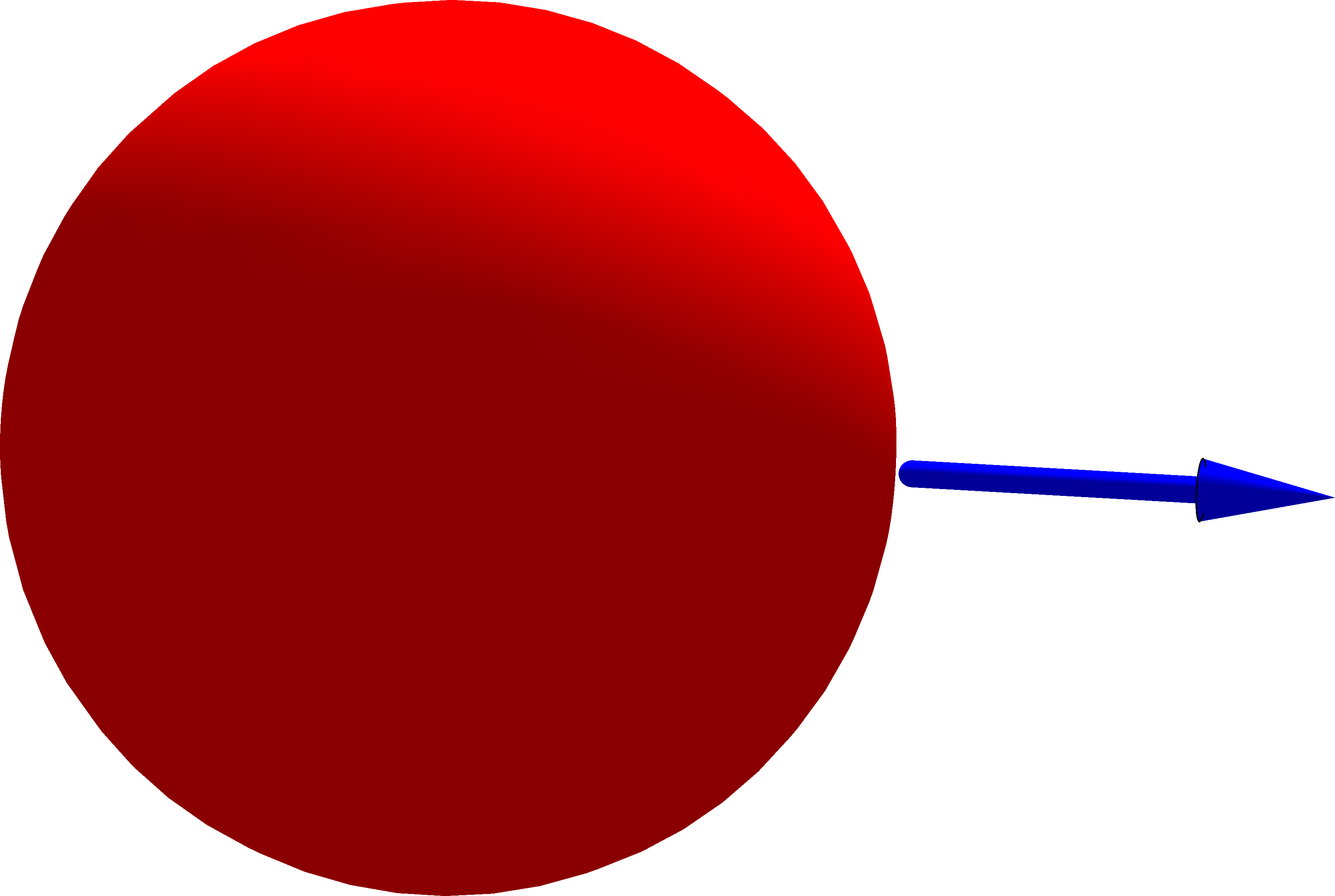}};
\node[name=c,below=1.5cm of a] {\includegraphics[scale=0.02]{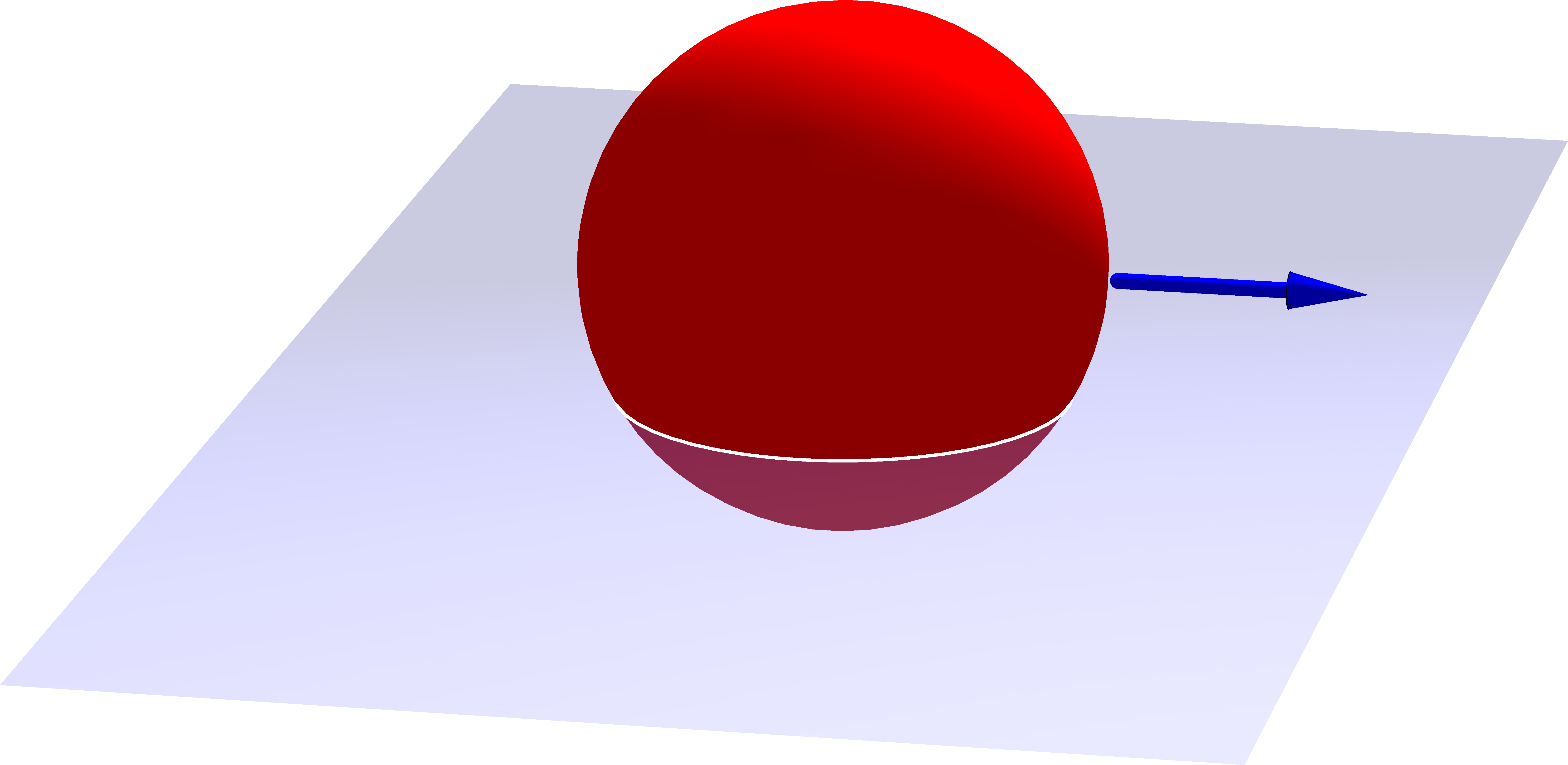}};
\node[name=d,below=1.5cm of b,yshift=-0.27cm] {\includegraphics[scale=0.02]{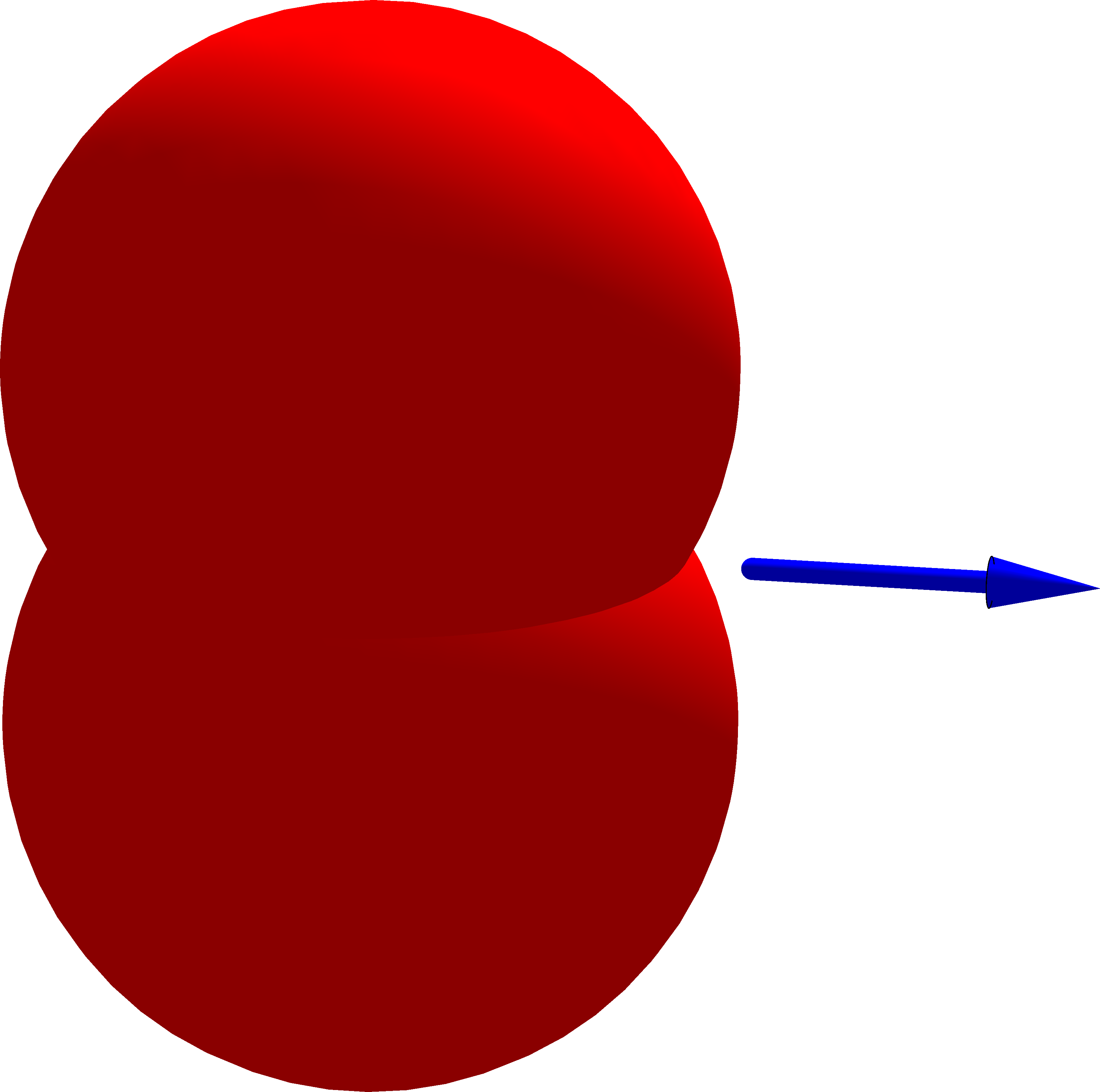}};
\node[name=e,right=of d,xshift=0.6cm,yshift=-0.21cm] {\includegraphics[scale=0.02]{Bild2b_500_klein}};
\node[text width=10ex] at ($(b)+0.9*(1,-1)$) {$\vek u=\umit$\newline$\vek T=\Tmit$};
\node[text width=10ex] at ($(e)+0.9*(1,-1)$) {$\vek u=\uohne$\newline$\vek T=\Tohne$};
\node at ($(a)+1*(-2,0.75)$) {(a)};
\node at ($(c)+1*(-2,1)$) {(b)};
\node[rotate=-3.2,color=white!30!black,yshift=0.07cm] at ($(a)+1*(-3.2:2.6)$) {\scalebox{2}{\ding{225}}};
\node[rotate=-3.2,color=white!30!black,yshift=-0.02cm] at ($(c)+1*(-3.2:2.65)$) {\scalebox{2}{\ding{225}}};
\node[rotate=-3.2,color=white!30!black,yshift=-0.05cm] at ($(d)+1*(-3.2:2)$) {\scalebox{2}{\ding{225}}};
\end{tikzpicture}
\caption{Pair of flow problems connected by the Lorentz reciprocal theorem~\eqref{eq:Lorentz}: (a) a sphere moving along a fluid-fluid interface with vanishing viscosity ratio causes a flow field identical to the classical Stokes flow problem in a bulk fluid, denoted by~$\umit$ and~$\Tmit$, if the contact angle~$\Theta$ equals~90\textdegree; (b) for~$\Theta\neq90$\textdegree{}, the flow problem is equivalent to a pair of fused spheres moving through a bulk fluid; by means of a perturbation expansion, the boundary condition on the complex particle surface is projected onto a sphere; the flow field in this case is denoted by~$\uohne$ and~$\Tohne$.}%
\label{fig:FlowProblems}%
\end{figure}

Our modeling focuses on the drag coefficient of a rigid sphere translating along a fluid-fluid interface at low Reynolds number. We hereby study the fundamental case of a pure fluid-fluid interface. Therefore, we do not employ any incompressibility constraint for the interface and assume a vanishing interfacial viscosity. Also, we neglect any deformation of the fluid-fluid interface on the scale of the particle radius~$a$, corresponding to a negligible influence of external forces acting normal to the fluid-fluid interface, such as buoyancy or electromagnetic forces. Assuming a planar interface also implies that the capillary number~$\Ca\definiert\mu_1 U / \sigma$ (with~$U=\|\vek U\|$ and the fluid-fluid interfacial tension~$\sigma$), measuring dynamic deformations of the fluid-fluid interface, is small compared to unity \citep{Radoev1992}. In addition to interfacial deformations, we neglect particle rotation. The validity of this assumption depends on the conditions at the three-phase contact line, for which two limiting cases exist. Firstly, the contact line can be pinned to defects of the particle surface, meaning that it retains a fixed position with respect to the latter. A pinned contact line prevents the particle from rotating if the capillary forces due to the fluid-fluid interface are sufficiently large. Secondly, the contact line can move tangentially to the particle surface if the contact-angle hysteresis is small or vanishes. In this case, the particle may rotate with an angular velocity~$\Omega$ dependent on the rate of dissipation occurring at the contact line. If the angular velocity is negligible compared to~$U/a$, the particle may be approximately considered non-rotating, because then the particle's surface velocity associated with the rotational motion is much smaller than the surface velocity~$U$ associated with the translational motion. For increasing angular velocity, that is, decreasing dissipation at the moving contact line, the assumption of a non-rotating particle looses its validity. Consequently, the following model is valid for a particle with~$\Omega a/U\ll1$. To further simplify the mathematical treatment, we assume a vanishing viscosity ratio between the two fluids,~$\mu_2/\mu_1\to0$. Thus the planar fluid-fluid interface effectively becomes a symmetry plane from the viewpoint of fluid~1, while the influence of fluid~2 can be neglected. For this reason, the flow problem is analogous to the motion of a body possessing reflection symmetry moving parallel to its symmetry plane. Clearly, for a contact angle~$\Theta$ of~90$^\circ$, the symmetric body is spherical, resulting in the classical Stokes flow problem around a sphere. The corresponding drag on an interfacial particle is then simply half the Stokes drag in the bulk of fluid~1 \citep{Radoev1992,Danov1995,Petkov1995,Ally2010}, implying
\begin{equation}
f(90^\circ,0)=1/2,
\label{eq:f90}
\end{equation}
where the function~$f$ is defined by \gleich{eq:StokesEinstein}. For contact angles differing from 90$^\circ$, the symmetric body consists of two fused spheres (cf. figure~\ref{fig:FlowProblems}(b)). This case has been studied by \citet{Zabarankin2007}, who provides drag coefficient values derived from a numerical solution of a Fredholm integral equation. Recently, \citet{Doerr2014} have derived the asymptotic expression
\begin{equation}
f(\Theta,0)=\frac{1}{2}\left[1+\frac{9}{16}\cos\Theta+\mathcal{O}(\cos^2\Theta)\right].
\label{eq:fDoerr1}
\end{equation}
The result~\eqref{eq:fDoerr1} has been obtained following a method by \citet{Brenner1964} (cf. \citet{Doerr2014} concerning necessary corrections to the method), which is based on spherical harmonics expansions and yields the velocity and pressure fields around a slightly deformed sphere. To this end, the particle shape (given by the pair of fused spheres of radius~$a$ in our case) needs to be parameterised in spherical coordinates~$(r,\theta,\varphi)$ according to
\begin{equation}
r=r_p(\theta,\varphi),
\label{eq:rthetaphi}
\end{equation}
and subsequently expanded in a power series 
\begin{equation}
r_p(\theta,\varphi)=a\left[1+\varepsilon\phi\ord1(\theta,\varphi)+\varepsilon^2\phi\ord2(\theta,\varphi)+\dotsb\right]
\label{eq:rentwallg}
\end{equation}
in terms of a small parameter~$\varepsilon$. Here, we choose~$\varepsilon=\cos\Theta$, so that~$2\varepsilon a$ equals the distance between the centers of the fused spheres. Accordingly, if we assume the centers of the spheres to lie on the $x$-axis with the symmetry plane given by~$x=0$, the particle shape is described by
\begin{equation}
r_p(\theta,\varphi)=a\left[1+\varepsilon\sin\theta\left|\cos\varphi\right|+\varepsilon^2\frac{\sin^2\theta\cos^2\varphi-1}{2}+\dotsb\right],
\label{eq:rentwspez}
\end{equation}
from which the functions~$\phi\ord1$ and~$\phi\ord2$ in \gleich{eq:rentwallg} can be read off. At the same time, the velocity and pressure fields, $\vek u$ and~$p$, are written in the form
\begin{align}
\vek u&=\vek u\ord0+\varepsilon\vek u\ord1+\varepsilon^2\vek u\ord2+\landaueps3\label{eq:uentw},~\text{and}\\
\vek p&=p\ord0+\varepsilon p\ord1+\varepsilon^2p\ord2+\landaueps3.\label{eq:pentw}
\end{align}
The particle moves with the velocity~$\vek U=U\vek e_z$. Therefore, the flow field obeys the boundary conditions
\begin{equation}
\left.\vek u\right|_{\Sigma_p}=\vek U
\label{eq:USigmap}
\end{equation}
at the particle surface~$\Sigma_p$, and
\begin{equation}
\left.\vek u\right|_{\Sigma_\infty}=0
\label{eq:USigmainf}
\end{equation}
on a spherical surface~$\Sigma_\infty$ at~$r\to\infty$. While condition~\eqref{eq:USigmainf} is readily adapted to the perturbation expansion~\eqref{eq:uentw}, condition~\eqref{eq:USigmap} on the particle surface requires a Taylor series expansion for removal of the implicit dependence on the shape parameter~$\varepsilon$. To be precise, the boundary condition~\eqref{eq:USigmap} in conjunction with the expanded particle shape~\eqref{eq:rentwallg} reads
\begin{equation}
\vek u(r_p,\theta,\varphi)=\vek U,
\label{eq:urp}
\end{equation}
which equals
\begin{equation}
\vek u\ord0(r_p,\theta,\varphi)+\varepsilon\vek u\ord1(r_p,\theta,\varphi)+\varepsilon^2\vek u\ord2(r_p,\theta,\varphi)+\landaueps3=\vek U.
\label{eq:uRB1}
\end{equation}
In \gleich{eq:uRB1}, the argument~$r_p$ depends on~$\varepsilon$, so that the velocity vectors~$\vek u\ord i~(i\in\{0,1,2\})$ are to be expanded in a Taylor series about~$r=a$ in powers of~$\varepsilon$, using equation~\eqref{eq:rentwallg}. After performing the expansions, inserting the results into equation~\eqref{eq:uRB1}, and grouping of terms, we arrive at
\begin{equation}
\left.
\begin{aligned}
\vek u\ord0&= \vek U\\
\vek u\ord1&= -a \phi\ord1\frac{\partial\vek u\ord0}{\partial r}\\
\vek u\ord2&= -a \phi\ord2\frac{\partial\vek u\ord0}{\partial r}-\frac{a^2}{2} \left[\phi\ord1\right]^2\frac{\partial^2\vek u\ord0}{\partial r^2}-a \phi\ord1\frac{\partial\vek u\ord1}{\partial r}
\end{aligned}\right\}
\text{at~}r=a.
\label{eq:RBord}
\end{equation} 
Clearly, the zeroth order problem corresponds to a spherical particle moving with constant velocity~$\vek U$ in an unbounded fluid, for which the velocity and pressure fields 
\begin{align}
\vek u\ord0&=-\frac{1}{2}U\left(\frac{a}{r}\right)^2\left(\frac{a}{r}-3\frac{r}{a}\right)\cos\theta\vek e_r-\frac{1}{4}U\left(\frac{a}{r}\right)^2\left(\frac{a}{r}+3\frac{r}{a}\right)\sin\theta\vek e_\theta
\label{eq:u0},~\text{and}\\
p\ord0&=\frac{3}{2}\mu_1 Ua\frac{\cos\theta}{r^2}\label{eq:p0}
\end{align}
are well known \citep{Happel1983}. The first-order problem has been solved by \citet{Doerr2014} in the particle's rest frame. Because the frame of reference only affects the zeroth-order flow~$\vek u\ord0$ by addition or subtraction of the velocity field~$U\vek e_z$, the first-order flow field~$\vek u\ord1$ considered by \citet{Doerr2014} may be directly used in the present study. The supplementary material contains the complete set of expressions required to calculate the velocity field~$\vek u\ord1$. Since~$\vek u\ord1$ is equal to the infinite series~$\sum_{k=0}^{\infty}\vek u_k\ord1$ \citep{Brenner1964}, the number of included terms needs to be limited in practical calculations. The values reported below as well as in the supplementary information correspond to~$k\leq20$. With this choice, the $\landaueps2$-contribution to the drag coefficient can be calculated to three significant digits, as will be shown in the following section.

\section{Applying the Lorentz Reciprocity Theorem}

According to the above discussion, we are in a position to compute the velocity fields~$\vek u\ord i(a,\theta,\varphi),~i\in\{0,1,2\}$, on a spherical surface by means of equation~\eqref{eq:RBord}. In other words, we are faced with two flow problems involving a sphere of radius~$a$ in an unbounded fluid, see \bild{fig:FlowProblems}.
The first of these problems (\bild{fig:FlowProblems}(a)), constructed by setting~$\varepsilon$ to zero, consists of a sphere with a no-slip surface condition and translating with velocity~$U\vek e_z$. We shall denote the velocity and stress tensor fields belonging to this first problem by~$\umit$ and~$\Tmit$, respectively. The second problem (\bild{fig:FlowProblems}(b)), associated with the truncated perturbation expansion in~$\varepsilon$ and denoted by~$\uohne$ and~$\Tohne$, is given by a sphere with a prescribed surface velocity field of
\begin{equation}
\uohne(a,\theta,\varphi)=\sum_{i=0}^{\infty}\varepsilon^i\vek u\ord i(a,\theta,\varphi)=U\vek e_z+\varepsilon\vek u\ord1(a,\theta,\varphi)+\varepsilon^2\vek u\ord2(a,\theta,\varphi)+\landaueps3
\label{eq:RBohnedach},
\end{equation}
according to \gleich{eq:RBord}. Since the solutions~$(\umit,\Tmit)$ and~$(\uohne,\Tohne)$ correspond to the same flow geometry, they are related by the Lorentz reciprocal theorem for Stokes flow \citep{Lorentz1896,Happel1983},
\begin{equation}
\int\limits_\Sigma\left(\smash{\Tmit}\cdot\vek n\right)\cdot\uohne\,\mathrm{d}\Sigma=\int\limits_\Sigma\left(\Tohne\cdot\vek n\right)\cdot\umit\,\mathrm{d}\Sigma,
\label{eq:Lorentz}
\end{equation}
where~$\Sigma$ comprises the particle surface~$\Sigma_p$ (outer normal vector~$\vek n=-\vek e_r$) and the surface~$\Sigma_\infty$ at infinity (outer normal vector~$\vek n=\vek e_r$). The Lorentz theorem~\eqref{eq:Lorentz} has been exploited in a number of cases, for instance by \citet{Stone1996}, \citet{Masoud2014}, and \citet{Schoenecker2014}. Its use in the present study is inspired by the argumentation of \citet{Stone1996}. The contribution from the integral over~$\Sigma_\infty$ in \gleich{eq:Lorentz} vanishes because
\begin{equation}
\left\|\umit\right\|\sim r^{-1},~\left\|\uohne\right\|\sim r^{-1},~\left\|\Tmit\cdot\vek e_r\right\|\sim r^{-2},~\text{and~}\left\|\Tohne\cdot\vek e_r\right\|\sim r^{-2}~\text{for}~r\to\infty.
\label{eq:asym}
\end{equation}
The integral in the Lorentz theorem~\eqref{eq:Lorentz} thus reduces to an integral over the particle surface~$\Sigma_p$. Recalling that~$\Tmit\cdot\vek n=-\Tmit\cdot\vek e_r=3\mu_1U/(2a)\vek e_z$ \citep{Stone1996} and using equation~\eqref{eq:RBohnedach}, the Lorentz theorem can be written as
\begin{equation}
\int\limits_{\Sigma_p}\frac{3\mu_1}{2a}U\vek e_z\cdot\left[U\vek e_z+\varepsilon\vek u\ord1+\varepsilon^2\vek u\ord2+\landaueps3\right]\mathrm{d}\Sigma=\int\limits_{\Sigma_p}\left(\Tohne\cdot\vek n\right)\cdot U\vek e_z\mathrm{d}\Sigma.
\label{eq:int1}
\end{equation}
Since~$U\vek e_z$ is a constant vector and~$\int_{\Sigma_{p}}\mathrm{d}\Sigma=4\pi a^2$, \gleich{eq:int1} simplifies to
\begin{equation}
6\pi\mu_1U+\frac{3\mu_1}{2a}\vek e_z\cdot\int\limits_{\Sigma_p}\left[\varepsilon\vek u\ord1+\varepsilon^2\vek u\ord2+\landaueps3\right]\mathrm{d}\Sigma=\vek e_z\cdot \underbrace{\int\limits_{\Sigma_p}\Tohne\cdot\vek n\mathrm{d}\Sigma}_{=-\vek F_D}
\label{eq:int2}.
\end{equation}
The integral on the right-hand-side of \gleich{eq:int2} has been identified with the negative of the Stokes drag~$\vek F_D$ on the particle because~$\vek n=-\vek e_r$. Using equations~\eqref{eq:RBord},~\eqref{eq:u0},~\eqref{eq:rentwallg}, and~\eqref{eq:rentwspez} in conjunction with the velocity field~$\vek u\ord1$ according to \citet{Doerr2014}, the integral occurring in \gleich{eq:int2} can be evaluated, yielding
\begin{equation}
\vek F_D\cdot\vek e_z=-6\pi\mu_1Ua\left[1+\frac{9}{16}\varepsilon-0.139\varepsilon^2+\landaueps3\right].
\label{eq:FD}
\end{equation}
From the symmetry of the problem, it clearly follows that~$\vek F_D=F_D\vek e_z$, so that the drag coefficient~\eqref{eq:fdef} for the original problem of a spherical particle diffusing along a fluid-fluid interface of zero viscosity ratio is given by
\begin{equation}
f(\Theta,0)=\frac{1}{2}\left[1+\frac{9}{16}\cos\Theta-0.139\cos^2\Theta+\mathcal{O}(\cos^3\Theta)\right].
\label{eq:fDoerr2}
\end{equation}
Correspondingly, for the diffusion coefficient~\eqref{eq:StokesEinstein} it follows that
\begin{equation}
D=\frac{16 kT}{3\pi\mu_1 a \left[16+9\cos\Theta+2.224\cos^2\Theta+\mathcal{O}(\cos^3\Theta)\right]}.
\label{eq:D}
\end{equation}
Note that the numerical coefficient~0.139 in \gleich{eq:fDoerr2} can be calculated to any desired number of significant digits, provided that a sufficient number of terms is considered in the spherical harmonics expansion by \citet{Brenner1964} and~\citet{Doerr2014}. As stated above, the value $0.139...$ corresponds to a number of~20 terms. The underlying rational expression reads~$765\,368\,413\,099/5\,497\,558\,138\,880$.
\section{Discussion}
In \bild{fig:Vergleich}, we compare the result~\eqref{eq:fDoerr2} and its first-order part~\eqref{eq:fDoerr1} to experimental and theoretical drag coefficient values from the literature.
\begin{figure}
\centering
\begin{tikzpicture}%
\begin{axis}[legend style={font=\footnotesize,cells={anchor=west},at={(axis cs:90,1)},anchor=north east,outer sep=0pt,inner xsep=2pt},line width={0.8pt},
	xlabel={$\Theta$ in \textdegree},
	ylabel={$f$},
	width=0.8\textwidth,
	height=0.618*0.8\textwidth,
	xmin=0,
	xmax=90,
	ymin=0.5,
	ymax=1,
	xtick={0,10,20,30,40,50,60,70,80,90},
	ytick={0.5,0.55,0.6,0.65,0.7,0.75,0.8,0.85,0.9,0.95,1}]
\addplot[color=violet,dash pattern=on 5pt off 1pt on 1pt off 1pt] file {f_von_Theta.tex_daten_1.dat};\addlegendentry{Equation~\eqref{eq:fDoerr1}}
\addplot[color=red] file {f_von_Theta.tex_daten_2.dat};\addlegendentry{Equation~\eqref{eq:fDoerr2}}
\addplot[color=teal,only marks,mark=o] file {f_von_Theta.tex_daten_3.dat};\addlegendentry{\citet{Zabarankin2007}}
\addplot[color=orange!80!red,only marks,mark=square] file {f_von_Theta.tex_daten_4.dat};\addlegendentry{\citet{Petkov1995}}
\addplot[color=olive,only marks,mark=otimes] file {f_von_Theta.tex_daten_5.dat};\addlegendentry{\citet{Danov1995}}
\addplot[color=blue,densely dashed] file {f_von_Theta.tex_daten_6.dat};\addlegendentry{\citet{Wang2013}}
\end{axis}
\end{tikzpicture}%
\caption{Comparison of the models~\eqref{eq:fDoerr1} and~\eqref{eq:fDoerr2} with experimental and theoretical drag coefficient values taken from the literature. In the experiments by \citet{Petkov1995}, an air-water interface with a very small viscosity ratio, $\mu_2/\mu_1\approx0.02$, was studied. The remaining curves are valid under the assumption~$\mu_2/\mu_1=0$.}%
\label{fig:Vergleich}%
\end{figure}
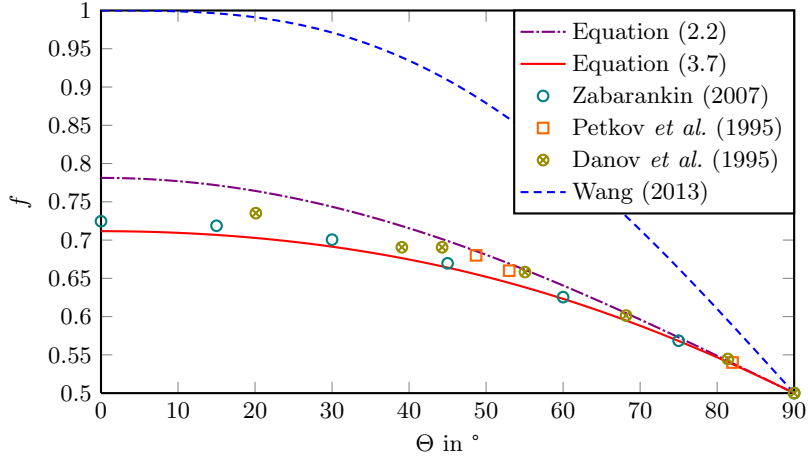
The numerical solution to the full flow problem by \citet{Zabarankin2007} may serve as a reference. Remarkably, \gleich{eq:fDoerr1} can be seen to be applicable at least down to a contact angle of~60\textdegree, where it still agrees well with the reference data. The inclusion of the $\cos^2\Theta$-term, resulting in \gleich{eq:fDoerr2}, leads to a significantly better agreement between the asymptotic model and the numerical solution. Within a maximum error in the drag coefficient~$f$ of less than~0.02 (3\%), the quadratic expression may even be applied to the full range of contact angles between~0 and~90\textdegree. An approximate formula by \citet{Wang2013},
\begin{equation}
f_W=1+\frac{\sin(2\Theta)-2\Theta}{2\pi},
\label{eq:Wang}
\end{equation}
based on the particle's cross-sectional area seen by the respective fluid phases is also plotted in \bild{fig:Vergleich}. Apparently, this expression is at variance with the other four data sets shown in the figure. The region where~$\Theta>90$\textdegree{} is excluded from \bild{fig:Vergleich} owing to a lack of reference data. The drag coefficient is expected to approach zero when the contact angle approaches 180\textdegree{}, a behaviour which is not reproduced by the model equation~\eqref{eq:fDoerr2}. Therefore, the range of validity of the model for contact angles larger than~90\textdegree{} remains an open issue. In order to model the drag coefficient near~180\textdegree{}, either the above perturbation expansion would need to be extended to higher orders in~$\varepsilon$, or a reference body differing from a sphere (such as a disk or an ellipsoid) would need to be used.

\section{Conclusions}
With equations~\eqref{eq:fDoerr2} and~\eqref{eq:D}, we have developed explicit expressions for the drag and diffusion coefficients of a spherical particle attached to the interface between two immiscible fluids for the case of a small viscosity ratio between the two phases. The relations account for the dependence on the contact angle between the two fluids and the solid surface. Following from the assumption on the viscosity ratio, the drag and diffusion coefficients of a pair of fused spheres moving perpendicular to their line-of-centers has been found simultaneously. A comparison between the model and reference data has shown that the model can be applied to the entire range of contact angles below~90\textdegree{} with high accuracy. By applying the Lorentz reciprocal theorem to a geometric perturbation expansion, one order of approximation has been gained for the drag and diffusion coefficients as compared to the flow field. In other words, the second-order result~\eqref{eq:fDoerr2} is based on the first-order velocity field, while calculating the non-trivial first-order result~\eqref{eq:fDoerr1} only requires the well-known flow field around a sphere. The method can be applied to any particle shape resulting from a small geometric modification of another particle shape with a known flow field. 

\section*{Acknowledgments}
We gratefully acknowledge financial support by the German Research Foundation through Grant No. HA 2696/25-1.

\bibliographystyle{jfm}
\bibliography{Literature_Diffusion_Coefficient}

\begin{appendix}
\parindent0pt
\newcommand{\platz}{,$ $}
\newcommand{\linie}{\rule[-1ex]{\linewidth}{0.8pt}}

\section{First-order velocity field}
\subsection{Series representation of the velocity field}
The first-order velocity field~$\vek u\ord1$ can be calculated according to the method by \citet{Brenner1964}. However, as pointed out by \citet{Doerr2014}, corrections to Brenner's method are necessary. Here, we provide all of the expressions required to explicitly compute the velocity field~$\vek u\ord1$, which  is of the form \citep{Brenner1964}
\begin{equation}
\vek u\ord1=\sum_{k=0}^{\infty}\vek u_k\ord1.
\label{eq:usum1}
\end{equation}
To facilitate the explicit evaluation of the velocity field and to simultaneously ensure a high degree of accuracy, we truncate the series after~20 summation terms and arrive at
\begin{equation}
\vek u\ord1\approx\sum_{k=0}^{20}\vek u_k\ord1.
\label{eq:usum2}
\end{equation}
The velocity fields~$\vek u_k\ord1$ can be expressed by
\begin{equation}
\begin{split}
\vek u_k\ord1=&\sum_{n=1}^{\infty}\left[\nabla\times\left(\vek r \kchin\right)+\nabla\left(\kphin\right)-\frac{n-2}{2n(2n-1)\mu_1}r^2\nabla\left(\kpn\right)\right.\\ &\left.+\vek r\frac{n+1}{n(2n-1)\mu_1}\kpn \right],
\end{split}
\label{eq:uk}
\end{equation}
where
\begin{align}
\kpn&= \frac{(2n-1)\mu_1}{(n+1)a}\left(\frac{a}{r}\right)^{n+1}\left[(n+2)\kXn+\kYn\right],\\
\kphin&=\frac{a}{2(n+1)}\left(\frac{a}{r}\right)^{n+1}\left(n \kXn+\kYn\right) ,~\text{and}\\
\kchin&=\frac{1}{n(n+1)}\left(\frac{a}{r}\right)^{n+1}\kZn
\end{align}
\citep{Brenner1964}. Through accommodating the velocity field~\eqref{eq:uk} to the boundary condition at the particle surface, the functions~$\kXn$,~$\kYn$ and~$\kZn$ can be specified. One has
\begin{align}
\kXn&=0 \quad\text{for}~n\leq1,\label{eq:kXn}\\
\kYn&=
\begin{cases}
\dfrac{3}{2}\displaystyle\sum_{m=0}^{k}\dfrac{c_{km}N_{km}\cos(m\varphi)}{2k+1}(k-1)(k+m)P_{k-1}^m(\cos\theta) & \text{for}~n=k-1\\
-\dfrac{3}{2}\displaystyle\sum_{m=0}^{k}\dfrac{c_{km}N_{km}\cos(m\varphi)}{2k+1}(2+k)(k-m+1)P_{k+1}^m(\cos\theta) & \text{for}~n=k+1\\
0 & \text{for all other}~n,\end{cases}\label{eq:kYn}\\
\intertext{and}
\kZn&=\begin{cases}
\frac{3}{2}U\sum_{m=0}^{k}mc_{km}N_{km}P_k^m(\cos\theta)\sin(m\varphi)\quad\text{for}~n=k\\
0\quad\text{for}~n\neq k.
\end{cases}
\label{eq:kZn}
\end{align}
In equations~\eqref{eq:kXn}--\eqref{eq:kZn}, the $P_k^m$ are associated Legendre polynomials. Expression~\eqref{eq:kYn}, developed by \citet{Doerr2014}, differs from the corresponding result by \citet{Brenner1964}, while equations~\eqref{eq:kXn} and~\eqref{eq:kZn} have been adopted from \citet{Brenner1964} without modification. The expansion coefficients~$c_{km}$ and $N_{km}$ are given by
\begin{equation}
N_{km}=
\begin{cases}
\sqrt{2} \sqrt{\frac{2k+1}{4\pi}\frac{(k-m)!}{(k+m)!}} &~\text{if}~m>0\\
\sqrt{\frac{2k+1}{4\pi}} &~\text{if}~m=0\\
\end{cases}
\label{eq:Nkm}
\end{equation}
and
\begin{equation}
c_{km}=\int \limits_{0}^{\pi} \int \limits_{0}^{2\pi} \sin\theta\left|\cos\varphi\right| N_{km} P_k^m(\cos\theta)\cos(m\varphi)\sin\theta\mathrm{d}\theta\mathrm{d}\varphi,
\label{eq:ckm}
\end{equation}
respectively. Equation~\eqref{eq:ckm} contains information about the particle geometry via the term~$\sin\theta\left|\cos\varphi\right|$, which is equal to the first-order term~$\phi\ord1$ in the expansion of the particle shape (equation~(2.5) in the paper). The velocity field~$\vek u\ord1$ according to \gleich{eq:usum1} is thus fully determined. For the sake of convenience, we explicitly display the coefficients~$c_{km}$ for $k\leq20$ in appendix~\ref{app:ckm}. Due to the symmetry of the particle shape with respect to the plane~$x=0$, coefficients with~$m<0$ and/or~$m$ odd vanish.

\subsection{Coefficients $\boldsymbol{c_{km}}$}\label{app:ckm}

\lineskip1ex
\subsubsection*{k=0: m=0}
$
 \sqrt{\pi } $
	\subsubsection*{k=2: m=0,2}
	$
 -\frac{\sqrt{5 \pi }}{8} \platz \frac{\sqrt{15 \pi }}{8} $
	\subsubsection*{k=4: m=0,2,4}
	$
 -\frac{3 \sqrt{\pi }}{64} \platz \frac{\sqrt{5 \pi }}{32} \platz -\frac{\sqrt{35 \pi }}{64}
    $
	\subsubsection*{k=6: m=0,2,4,6}
	$
 -\frac{5 \sqrt{13 \pi }}{1024} \platz \frac{\sqrt{\frac{1365 \pi }{2}}}{1024} \platz
   -\frac{3 \sqrt{91 \pi }}{1024} \platz \frac{\sqrt{\frac{3003 \pi }{2}}}{1024}  $
	\subsubsection*{k=8: m=0,2,4,6,8}
	$
 -\frac{35 \sqrt{17 \pi }}{16384} \platz \frac{3 \sqrt{\frac{595 \pi }{2}}}{4096} \platz
   -\frac{3 \sqrt{1309 \pi }}{8192} \platz \frac{\sqrt{\frac{7293 \pi }{2}}}{4096} \platz
   -\frac{3 \sqrt{12155 \pi }}{16384} $
	\subsubsection*{k=10: m=0,2,\dots,10}
	$
 -\frac{147 \sqrt{21 \pi }}{131072} \platz \frac{49 \sqrt{385 \pi }}{131072} \platz -\frac{7
   \sqrt{5005 \pi }}{65536} \platz \frac{21 \sqrt{\frac{5005 \pi }{2}}}{131072} \platz
   -\frac{7 \sqrt{\frac{85085 \pi }{3}}}{131072} \platz \frac{7 \sqrt{\frac{323323 \pi
   }{6}}}{131072} $
	\subsubsection*{k=12: m=0,2,\dots,12}
	$
 -\frac{3465 \sqrt{\pi }}{1048576} \platz \frac{45 \sqrt{3003 \pi }}{524288} \platz
   -\frac{225 \sqrt{\frac{1001 \pi }{2}}}{1048576} \platz \frac{75 \sqrt{\frac{2431 \pi
   }{2}}}{524288} \platz -\frac{15 \sqrt{138567 \pi }}{1048576} \platz \frac{15
   \sqrt{\frac{88179 \pi }{2}}}{524288} \platz -\frac{15 \sqrt{\frac{676039 \pi
   }{2}}}{1048576}  $
	\subsubsection*{k=14: m=0,2,\dots,14}
	$
 -\frac{14157 \sqrt{29 \pi }}{33554432} \platz \frac{1089 \sqrt{39585 \pi }}{67108864} \platz
   -\frac{99 \sqrt{\frac{2467465 \pi }{2}}}{33554432} \platz \frac{33 \sqrt{46881835 \pi
   }}{67108864} \platz -\frac{33 \sqrt{12785955 \pi }}{33554432} \platz \frac{33
   \sqrt{58815393 \pi }}{67108864} ,\\ -\frac{165 \sqrt{\frac{1508087 \pi
   }{2}}}{33554432} \platz \frac{495 \sqrt{646323 \pi }}{67108864} $
	\subsubsection*{k=16: m=0,2,\dots,16}
	$
 -\frac{306735 \sqrt{33 \pi }}{1073741824} \platz \frac{20449 \sqrt{935 \pi
   }}{268435456} \platz -\frac{1573 \sqrt{\frac{323323 \pi }{2}}}{268435456} \platz
   \frac{3003 \sqrt{46189 \pi }}{268435456} ,\\ -\frac{1001 \sqrt{\frac{5311735 \pi
   }{3}}}{536870912} \platz \frac{715 \sqrt{\frac{2860165 \pi }{3}}}{268435456} \platz
   -\frac{2145 \sqrt{\frac{245157 \pi }{2}}}{268435456} \platz \frac{143 \sqrt{35547765
   \pi }}{268435456} \platz -\frac{143 \sqrt{1101980715 \pi }}{1073741824} $
	\subsubsection*{k=18: m=0,2,\dots,18}
	$
 -\frac{1738165 \sqrt{37 \pi }}{8589934592} \platz \frac{61347 \sqrt{59755 \pi
   }}{8589934592} \platz -\frac{5577 \sqrt{\frac{920227 \pi }{2}}}{2147483648} \platz
   \frac{1001 \sqrt{\frac{117920517 \pi }{2}}}{4294967296} \platz -\frac{15015
   \sqrt{274873 \pi }}{4294967296} ,\\ \frac{2145 \sqrt{\frac{28861665 \pi
   }{2}}}{4294967296} \platz -\frac{2145 \sqrt{\frac{7971317 \pi }{2}}}{2147483648} \platz
   \frac{429 \sqrt{\frac{3706662405 \pi }{2}}}{8589934592} \platz -\frac{429
   \sqrt{2398428615 \pi }}{8589934592} \platz \frac{715 \sqrt{\frac{3357800061 \pi
   }{2}}}{8589934592}  $
	\subsubsection*{k=20: m=0,2,\dots,20}
	$
 -\frac{10207769 \sqrt{41 \pi }}{68719476736} \platz \frac{48841 \sqrt{899745 \pi
   }}{34359738368} \platz -\frac{8619 \sqrt{117266765 \pi }}{68719476736} \platz \frac{1105
   \sqrt{\frac{914680767 \pi }{2}}}{17179869184} \platz -\frac{9945 \sqrt{23453353 \pi
   }}{34359738368} ,\\[0.5ex] \frac{1989 \sqrt{\frac{309157835 \pi }{2}}}{17179869184} \platz
   -\frac{3315 \sqrt{\frac{1916778577 \pi }{2}}}{68719476736} \platz \frac{3315
   \sqrt{\frac{531543639 \pi }{2}}}{34359738368} \platz -\frac{3315 \sqrt{1240268491 \pi
   }}{68719476736} \platz \frac{1105 \sqrt{\frac{7245779079 \pi }{2}}}{34359738368} ,\\[0.5ex]
   -\frac{663 \sqrt{\frac{156991880045 \pi }{2}}}{68719476736} $
\end{appendix}


\end{document}